\documentclass[twocolumn,notitlepage,prx,superscriptaddress,longbibliography]{revtex4-2}
\usepackage{amssymb}
\usepackage{algpseudocode}
\usepackage{algorithm}
\usepackage{amsmath}
\usepackage{braket}
\usepackage{booktabs}
\usepackage{color}
\usepackage{comment}
\usepackage{dsfont}
\usepackage{etoolbox}
\usepackage{epstopdf}
\usepackage[T1]{fontenc}
\usepackage{float}
\usepackage{graphicx}
\usepackage[colorlinks=true,linkcolor=blue,citecolor=blue,urlcolor=blue]{hyperref}
\usepackage[latin9]{inputenc}
\usepackage{indentfirst}
\usepackage{latexsym,bm,euscript}
\usepackage{mathrsfs}
\usepackage{multirow}
\usepackage{natbib}
\usepackage{soul}
\usepackage{subfigure}
\usepackage{times}
\usepackage{titlesec}
\usepackage{tikz}
\usepackage{txfonts}
\usepackage{xspace}
\usepackage{xfrac}
\usepackage{etoolbox}

\newcommand{\Tr}[1]{\mathrm{Tr}\left\{#1\right\}}

\begin{document}
\title{Detecting Confined and Deconfined Spinons in Dynamical Quantum Simulations}

\author{Qiaoyi Li}
\affiliation{School of Physics, Beihang University, Beijing 100191, China}
\affiliation{CAS Key Laboratory of Theoretical Physics, 
Institute of Theoretical Physics, \\Chinese Academy of Sciences, Beijing 100190, China}

\author{Jian Cui}
\email{jiancui@buaa.edu.cn}
\affiliation{School of Physics, Beihang University, Beijing 100191, China}

\author{Wei Li}
\email{w.li@itp.ac.cn}
\affiliation{CAS Key Laboratory of Theoretical Physics, 
Institute of Theoretical Physics, \\Chinese Academy of Sciences, Beijing 100190, China}
\affiliation{School of Physics, Beihang University, Beijing 100191, China}
\affiliation{CAS Center for Excellence in Topological 
Quantum Computation, University of Chinese Academy 
of Sciences, Beijng 100190, China}

\begin{abstract} 
Dynamical spin-structure factor (DSF) contains fingerprint information of collective
excitations in interacting quantum spin systems. In solid state experiments, DSF can 
be measured through neutron scatterings. However, it is in general challenging to 
compute the spectral properties accurately via many-body simulations. Currently, 
quantum simulation and computation constitute a thriving research field, which are 
believed to provide a very promising platform for simulating quantum many-body 
systems. In this work, we establish a link between the many-body dynamics and 
quantum simulations by studying the non-equilibrium DSF (nDSF) measured on 
direct product states, which are accessible in contemporary quantum simulators 
with Rydberg atoms, superconducting qubits, etc. Based on the many-body 
calculations of transverse field Ising chains, we find the nDSF can be used to 
sensitively probe the multi-spinon continua associated with the two-spinon creation 
and the spinon-antispinon process, etc. Moreover, we further demonstrate that the 
low-energy spinons can be confined --- forming spinon bound states --- under a finite
longitudinal field. Our results pave the way of quantum simulation and manipulation 
of fractional excitations in highly-entangled quantum many-body systems.   
\end{abstract}

\date{\today}

\maketitle

\section{Introduction} 
\begin{figure}[H]
\centering
\includegraphics[width = 0.95\linewidth]{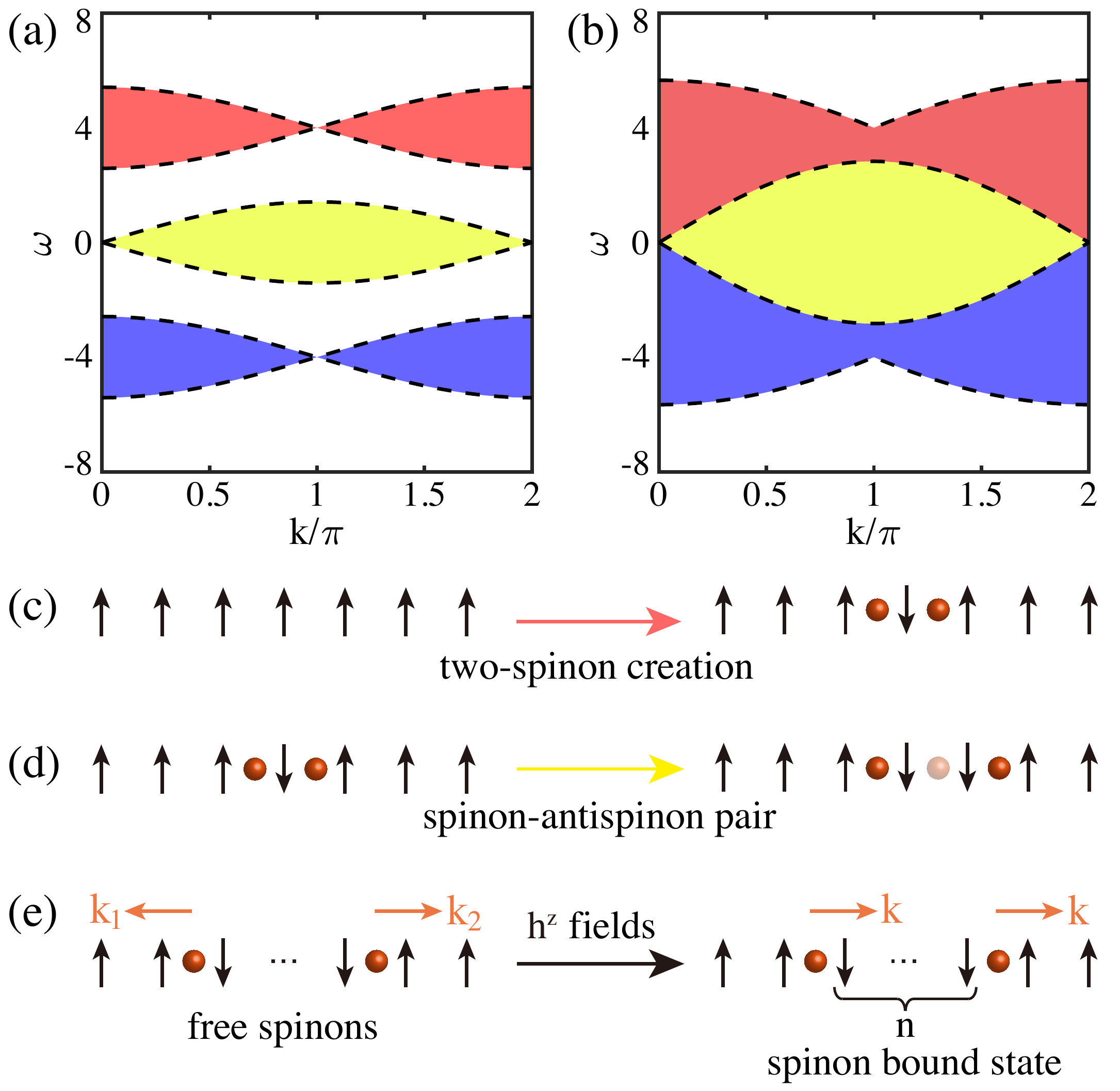}
\caption{(a) shows the bowtie- and shell-shape spin excitation 
continua present in the gapped phases of the TFI model, and (b) 
depicts the excitation continua right at the QCP, where the lower 
boundary of the bowtie excitations coincides with that of the upper 
boundary of the shell excitations. The cartoon pictures of the (c) 
two-spinon creation and (d) spinon-antispinon processes are illustrated, 
which are responsible for the bowtie and shell excitation continua. 
All these spinon continua are determined by $\omega(k) = \pm \epsilon(k_1) 
\pm \epsilon(k_2)$, where $k = k_1 + k_2$ and $\pm\epsilon(k)$ is 
the spinon ($+$) or antispinon ($-$) dispersion. The single-spinon 
dispersions used in the illustration are $\epsilon(k) \simeq 2 - 
\sin(2\gamma)\cos k$ (away from QCP) and $\epsilon(k) = 2\sqrt{2} 
\left|\sin\left(\frac{k}{2}\right)\right|$ (right at QCP). (e) Confinement of 
spinons under nonzero longitudinal fields, where the bound states consist
of two spinons at a distance $n = 1, 2,\cdots$ and move at the same quasi-momentum $k$.
}
\label{Fig1}    
\end{figure}

Hunting exotic quasi-particles in ultra quantum states of matter 
constitutes a fundamental research topic in modern condensed 
matter and quantum physics. One prominent example is the fractional 
spinon in one-dimensional (1D) spin chain, which is a charge 
neutral quasi-particle carrying fractional spin quantum number
\cite{Giamarchi2004Book}. Intuitively, the spinon excitation in a 
quantum Ising chain can be regarded as the domain wall moving 
freely in Fig.~\ref{Fig1}, which has been evidenced experimentally 
in spin-chain compounds via the inelastic neutron scattering 
measurements~\cite{Lake2013PRL,Mourigal2013NP,Wu2016Science}. 
Despite the existence of well-recognised model materials like the 
spin-1/2 Heisenberg chain compound KCuF$_3$~\cite{Lake2013PRL}, 
Ising-chain material CoNb$_2$O$_6$
\cite{Coldea2010,Morris2014,Fava2020,Morris2021}, the rare-earth 
Heisenberg XXZ chain Yb$_2$Pt$_2$Pb~\cite{Wu2016Science,
Gannon2019NC}, YbAlO$_3$~\cite{Wu2019NC}, etc, there inevitably 
exist additional interactions in solid-state compounds beyond 
the ``clean" models. It thus adds complexity in the analysis of 
the exotic excitations in these quantum magnets. 

On the other hand, simulations of the dynamical spin structure factor 
(DSF) are quite challenging in many-body calculations. Dynamical 
exact diagonalization suffers from strong finite-size effects, the quantum 
Monte Carlo methods have essential difficulties in dealing with real-time 
dynamics, and the matrix product state (MPS) based simulations do 
not have the access to long-time dynamics due to the rapid entanglement 
growth and thus the obtained spectral properties have rather limited 
resolution in frequencies. Therefore, to study the long-time dynamics 
of many-body systems we need to seek for different tools.

% ============ Fig. 1 ============ %
% =============================== %

{The rapid development of quantum technologies in past 
few years has witnessed the dawn of the quantum era
\cite{Ladd2010Nature,Georgescu2020Nature,Green2021Nature,
Leon2021Science}, and offers unique access to long-time 
evolution of quantum systems.} Although the universal quantum 
computing is still quite far from reach at the current moment, 
the idea of quantum simulation, i.e., emulating the problem 
to be solved via a highly controllable quantum system, 
is gradually becoming reality, which would fundamentally 
reformulate the routine that scientists conquer the uncharted 
territory of the realm of nature in the future~\cite{Georgescu2014}. 
In particular, physicists are making increasing endeavor to 
design quantum simulation protocols to solve many important 
problems in diverse fields of physics, ranging from  lattice gauge 
theory~\cite{QSLatticeGauge2019}, Kibble-Zurich mechanism
\cite{QSKZLukin2019}, many-body localization~\cite{MBLBloch2016,
MBLMonroe2016,MBLFan2018,MBLWangHH2021}, 
phase diagram of Hubbard model~\cite{Cheuk2016a,Cheuk2016b,
Mazurenko2017,Chiu2019,Koepsell2019} to quantum walk
\cite{QuantumWalk2019,QuantumWalk2021}, just to name a few. 

Therefore, to simulate dynamical quantities like the spectral 
functions of the many-body systems via quantum simulations 
constitutes a very appealing proposal. As the bottleneck of calculating 
DSF, being the long-time evolution of the quantum many-body 
system under study, is exactly the strength of quantum simulation, 
this proposal has a natural advantage. Nevertheless, there still exists 
\textit{an apparent gap} for the current quantum techniques 
to simulate the dynamical properties. One major limitation of 
current  experiments in common is that only a few states 
--- mostly product states and several special entangled states, 
such as the $W$ states~\cite{Browaeys2016WState} and the 
GHZ states~\cite{GHZLukin2019,GHZChina2019} --- can be 
prepared on quantum simulation experiments. 
Physically, it means only these special states can be used to 
probe the DSF of the system via quantum simulations, as opposed 
to the context of conventional condensed matter physics where
the ground states or thermal equilibrium states are adopted 
to gauge the excitation information of the system/phase. 
Given the fact that even though those entangled states 
can be prepared on certain quantum simulation platforms, 
the system size and lifetime are still very limited
\cite{Browaeys2016WState,GHZLukin2019,GHZChina2019}.

In this work, we fill this gap by systematically studying the 
quenched real-time evolution of quantum spin systems and 
computing the non-equilibrium DSF (nDSF) {for product initial 
states}. We consider the spin-1/2 transverse field Ising (TFI) chain 
which is readily accessible in contemporary quantum simulators, 
and find the product state serves as an excellent probing state 
in dynamical quantum simulations. In particular, we find the dynamical 
features of the deconfined spinons can be recognised in nDSF, which
contain the two-spinon creation/annihilation and spinon-antispinon 
continua, reflecting rich processes even beyond those reflected in their 
equilibrium counterpart. Prominent features in nDSF can also be employed 
to accurately pin down the quantum critical point (QCP), even when the 
ground state is absent. Furthermore, the spinons can be confined  by 
applying longitudinal fields, and they form two-spinon bound states at 
low while remain asymptotically free at higher energies. 

% ========== Fig. 2 ======== %
\begin{figure*}[ht]
    \centering
    \includegraphics[width = 0.9\linewidth]{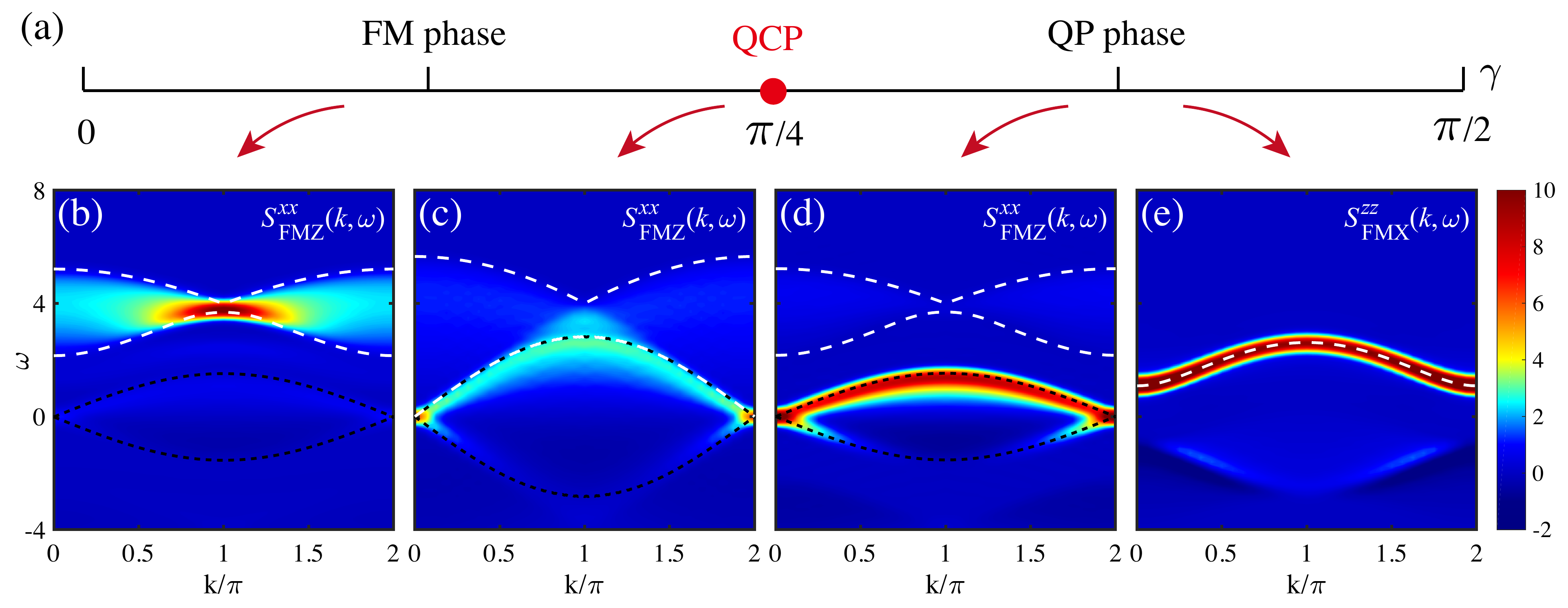}
    \caption{(a) represents a 1D phase diagram of the TFI chain, where a 
    QCP exists at $\gamma_c = \pi/4$ separating the FM and QP phases. 
    The nDSF $S^{xx}_{\rm FMZ}(k,\omega)$ results on the direct-product state
    $\ket{\Phi}_{\rm FMZ}$ along the $\sigma^z$ direction are measured 
    (b) in the FM phase ($\gamma = \pi/8$), (c) at QCP ($\gamma = \pi/4$), 
    and (d) in the QP phase ($\gamma = 3\pi/8$). The boundaries of two-spinon
    continua, determined from the exact spinon dispersion in the thermodynamic 
    limit (see Appendix Sec.~\ref{App:ExactTFI}), are labeled by white dashed lines, 
    and those of spinon-antispinon continua as black dotted lines. The boundaries 
    agree with our finite size calculations, up to the frequency resolution of the 
    spectral data. (e) is the longitudinal nDSF $S^{zz}_{\rm FMX}(k,\omega)$
    measured on the polarized state $\ket{\Phi}_{\rm FMX}$ along the 
    $\sigma^x$ direction, with the parameter $\gamma = 3\pi/8$ in the 
    Hamiltonian. The analytical dispersion of magnon is labeled by white 
    dashed line, also in excellent agreement with the numerical results. 
    }
    \label{Fig2}    
\end{figure*}
% ========================= %

\section{Model and Methods} 
Below we consider the TFI chain, a fundamental lattice model 
exhibiting fractional spinon excitations and quantum criticality 
driven by the transverse fields. The Hamiltonian of the TFI model 
reads,
\begin{equation}
H_{\rm TFI} = -\cos{\gamma}\sum_{i}\sigma_i^z\sigma_{i + 1}^z - 
\sin{\gamma}\sum_{i} \sigma_{i}^x,
\label{Eq:TFI}
\end{equation}
where $\sigma^\alpha$'s
are Pauli matrices and the parameter $\gamma$ controls 
the ration between the spin-spin couplings and transverse 
fields, and a QCP resides at exactly $\gamma_c = \pi/4$ 
[c.f., Fig.~\ref{Fig2}(a)] that separates the ferromagnetic (FM) 
phase ($0 \leq \gamma < \pi/4$) and quantum paramagnetic (QP) 
phase ($\pi/4 < \gamma \leq \pi/2$). The TFI model can be realized 
via Rydberg atoms~\cite{RydbergIsingBloch,Rydberg51Qubit,
GHZLukin2019} and therefore our many-body studies here also 
provide useful guide for observing the intriguing dynamical 
signatures of fractional spinon excitations in quantum simulations.

In the present study, we use the tensor network (TN) based real-time 
evolution approach~\cite{Vidal2004TEBD,White2004tDMRG,
Vidal2007iTEBD,Barthel2009} to compute the time-dependent 
spin correlations (see Appendix Sec.~\ref{App:TEBD}) and also 
the {nDSFs}. The key quantity computed with time-dependent 
TN method here --- and proposed to be measured in quantum 
simulations --- reads
\begin{equation}
\begin{aligned}
    {S^{\alpha\beta}(i, j, t_1, t_2)} \equiv& \left\langle
    \sigma_i^\alpha(t_1) \sigma_j^\beta(t_2)
    \right\rangle\\ =& \left\langle 
    \Phi\right| 
    e^{{\rm i}Ht_1}\sigma_i^\alpha e^{-{\rm i}H(t_1 - t_2)} \sigma_j^\beta 
    e^{-{\rm i}Ht_2} 
    \ket{\Phi},
\end{aligned}
\label{Eq:Correlation}
\end{equation}
where $\ket{\Phi}$ is an initial state hereafter chosen as a 
direct product state. Below we start from such easy-to-prepare 
state $\ket{\Phi}$ and compute the real-time evolution after 
quantum quenches, restricted to the case with $t_2=0$ that 
is also quite natural to measure in quantum simulations.
  
To be specific, as the time-dependent spin-spin correlation
$S^{\alpha\beta}(i, j, t_1, t_2=0)$ is a function of $t (\equiv t_1)$, 
we can then compute the nDSF
\begin{equation}
S^{\alpha\beta}(k, \omega) = \frac{1}{L}\sum_{i,j}e^{-{\rm i}k(i - j)}
\int_{-\infty}^{\infty} e^{{\rm i}\omega t} {S^{\alpha\beta}(i, j, t)} \, {\rm d}t 
\label{Eq:DSF}
\end{equation} 
to characterize the quenched spin dynamics on the initial 
state $\ket{\Phi}$ that is in general not the ground state or 
even eigenstates. We find below the nDSF in Eq.~(\ref{Eq:DSF}) 
can be used to extract fingerprint information of the spin 
excitations and are thus the major quantities of interest in 
the present work.

In the practical calculations, we use the time evolution block decimation 
approach~\cite{Vidal2004TEBD,White2004tDMRG,Vidal2007iTEBD} 
equipped with the folding trick~\cite{Banuls2009} to evolve the MPO 
representations of the $\sigma^\alpha$ operators from both sides 
(i.e., in the Heisenberg picture, see details in {Appendix Sec.~\ref{App:TEBD})}). The bond inversion symmetry and time 
reversal symmetry are employed to reduce computational costs 
and accelerate the simulations. The bond dimension used in the 
calculations is $D = 800$, which results in a relative truncation error 
$\lesssim 10^{-5}$ at the maximal evolution time $t_\textrm{M} = 12$. 
In order to improve the frequency resolution in computed nDSF, the {linear
prediction technique~\cite{White2008PRB,Barthel2009} has been} 
exploited to extrapolate the time-dependent correlation functions 
from $t_\textrm{M} = 12$ to $t_\textrm{M} = 18$. As $t_\textrm{M}$ 
is still rather limited, we employ the Parzen window function in the 
time-directional Fourier's transformation and relieve the artificial 
oscillations, which results in a frequency resolution with a standard 
error $\sigma_\omega = \sqrt{3}/9$.

\section{Results}
\subsection{Multi-Spinon Continua in nDSF}
In Fig.~\ref{Fig2}, we show our results of the nDSF $S(k,\omega)$ 
computed on different direct product states, which clearly show 
excitation continua that can be used to detect the fractional spinon 
excitations. To be specific, we consider below the nDSF 
$S^{xx}_{\rm FMZ}(k, \omega)$ of $\sigma^x$ operators 
measured in the FM state $\ket{\Phi}_{\rm{FMZ}} \equiv 
\ket{\uparrow \uparrow ... \uparrow}_z$ along the $\sigma^z$ 
direction, and $S^{zz}_{\rm FMX}(k, \omega)$ on the 
product state $\ket{\Phi}_{\rm{FMX}} \equiv \ket{\rightarrow 
\rightarrow ... \rightarrow}_x$ along the $\sigma^x$ direction,
respectively.  Intriguingly, from the nDSF results in Fig.~\ref{Fig2}(b-d)
we find excitation continua of the bowtie and shell-like shapes, 
which represent the two-spinon creation/annihilation and 
spinon-antispinon processes (c.f., Fig.~\ref{Fig1}), respectively.

For example, in Fig.~\ref{Fig2}(b) we see a clear bowtie excitation 
regime centered at a finite frequency, which represents the two-spinon 
continuum described by $\omega_{\rm bw}(k) = \epsilon(k_1) + \epsilon(k_2)$,
with $k=k_1+k_2$ and $\epsilon(k) = 2\sqrt{1 - \sin(2\gamma)\cos k}$ 
(See derivation of the spinon dispersion in Appendix Sec.~\ref{App:ExactTFI} 
and more discussions on the bowtie excitations in Appendix 
Sec.~\ref{App:Bowtie}). Such two-spinon creation process is 
illustrated in Fig.~\ref{Fig1}(c) with the corresponding bowtie 
continuum illustrated in Fig.~\ref{Fig1}(a). Moreover, in Fig.~\ref{Fig2}(b) 
we depict such upper and lower boundaries of $\omega_{\rm bw}(k)$ 
and find they bound very well the spinon continuum computed 
from the product state $\ket{\Phi}_{\rm{FMZ}}$, confirming that 
these excitation continua in nDSF indeed represent the two-spinon 
excitations. 

The spin flip process can not only create a pair of spinons 
--- the bowtie excitations  --- but also introduce a spinon-antispinon 
process  [c.f., Fig.~\ref{Fig1}(d)], as the initial state is not the ground 
state and can inherently contain spinon excitations already. It thus 
allows a spinon-antispinon process by a spin flip, and generates a 
shell-like excitation continua $\omega_{\rm sh}(k) = \epsilon(k_1) -
\epsilon(k_2)$, with the same single-spinon dispersion $\epsilon(k)$ 
used in deriving the bowtie excitations. In Fig.~\ref{Fig2}, we observe 
that the shell-like shape is rather faint in the FM phase in panel (b), 
as $\ket{\Phi}_{\rm{FMZ}}$ is rather close to the true ground state 
and there is no much spectral weight for such spinon-antispinon 
process. However, the shell-like excitations become very prominent 
in the QP phase with $\gamma>\gamma_c$ [c.f., Fig.~\ref{Fig2}(d)], 
where the product state $\ket{\Phi}_{\rm{FMZ}}$ is now far from 
the ground state and thus the spin flip process can annihilate a 
spinon and create another [c.f., Fig~\ref{Fig1}(d)]. At the same time, 
in Fig.~\ref{Fig2}(d) the two-spinon creation bowtie excitation continuum 
becomes virtually invisible. Interestingly, such a role swap of different 
multi-spinon processes in the prominent excitation continua features 
between Fig.~\ref{Fig2}(b) and (d) reflects the existence of a quantum 
phase transition controlled by the parameter $\gamma$.

Moreover, right at the QCP there emerge features in the system
in sharp distinction from both gapped FM and QP phases. At 
$\gamma_c=\pi/4$, the spinon dispersion is $\epsilon(k) = 
2\sqrt{2}\left|\sin\left(\frac{k}{2}\right)\right|$, which thus leads to 
$2\sqrt{2}\left|\sin\left(\frac{k}{2}\right)\right| \leq \omega(k) \equiv 
\epsilon(k_1) + \epsilon(k_2) \leq 4\sqrt{2}\left|\sin\left(\frac{k}{4}\right) 
\right|$ for $\pi<k<2\pi$, and symmetrically for $0<k<\pi$ around $k=\pi$. 
Besides, we have also plotted the upper and lower boundaries for the 
spin-antispinon shell regime, i.e., $\pm 2\sqrt{2} \left|\sin\left(\frac{k}{2}
\right)\right|$. Note that the upper boundary of spin-antispinon excitation 
coincides with the lower boundary of the two-spinon creation continuum (see Fig.~\ref{Fig1} 
and also Appendix Sec.~\ref{App:Bowtie}).

In Fig.~\ref{Fig1}, we have also illustrated the ``negative" two-spinon 
continua {corresponding to the annihilation process}. Such bowtie 
excitations with negative energies are not clearly seen in Fig.~\ref{Fig2} 
as the trial state $\ket{\Phi}$ is not ``far enough" from the ground state. 
In the Appendix Sec.~\ref{App:SubNegBowtie} we have provided an 
example where $\ket{\Phi}$ constitutes a very high energy state and 
such negative bowtie can be clearly observed there.

% =========== Fig. 3 ======== %
\begin{figure}[ht]
\centering
\includegraphics[width = 1\linewidth]{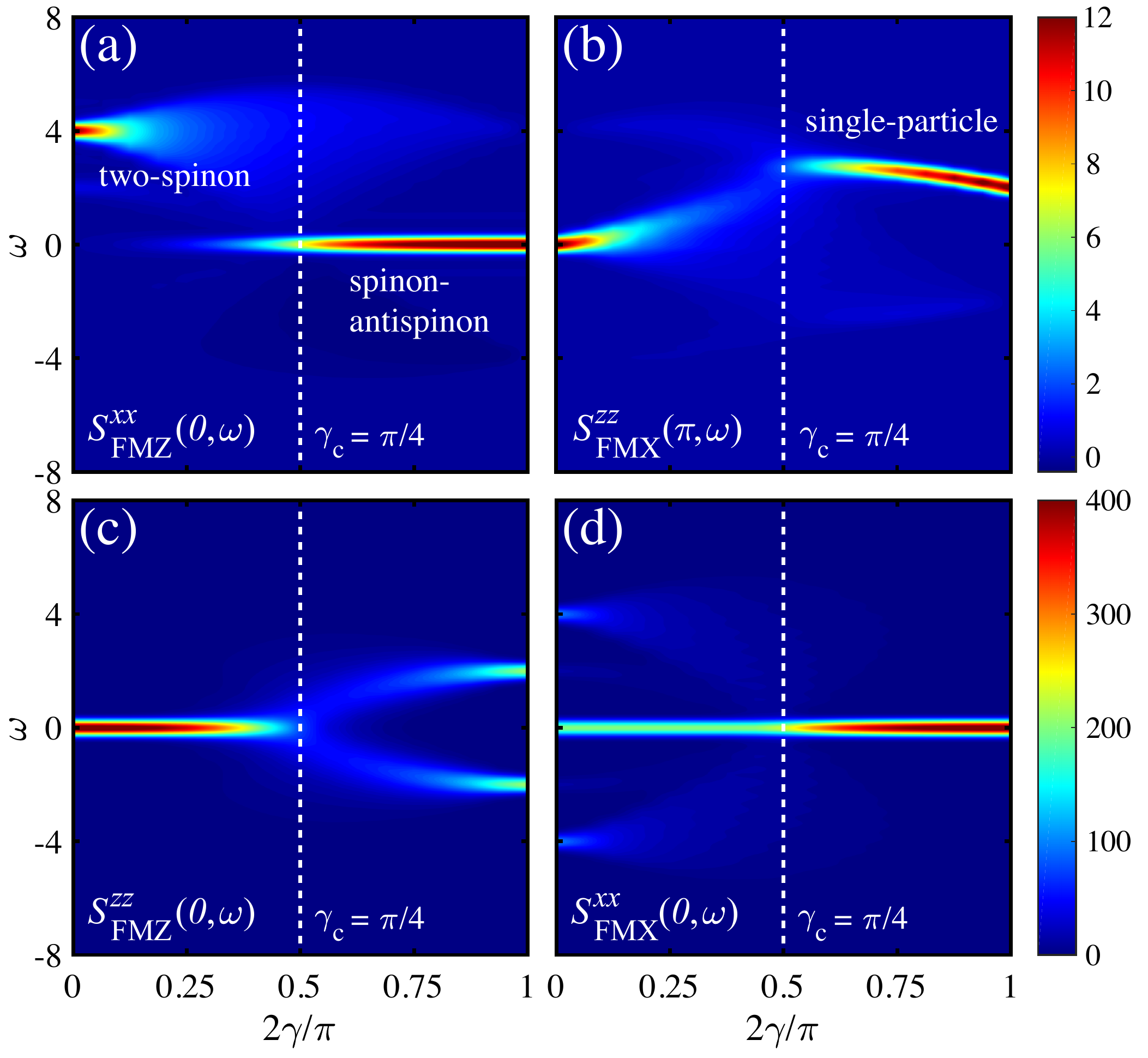}
\caption{The nDSF results computed on different direct product states 
$\ket{\Phi}$, scanned vs. various $\gamma$ values. (a, c) show the $k$-cut of
$S^{xx}_{\rm FMZ}(k,\omega)$ and $S^{zz}_{\rm FMZ}(k,\omega)$
computed on $\ket{\Phi}_{\rm FMZ}$ state and (b, d) are the two nDSF
results measured on the $\ket{\Phi}_{\rm FMX}$ state.
}
\label{Fig3}    
\end{figure}

\subsection{Lehmann's Spectral Representation of the nDSF}
The results in Fig.~\ref{Fig2} show that the nDSF measured on 
$\ket{\Phi}$ can reflect the multi-spinon continua quantitatively, and 
besides provide unique access to the spinon-antispinon process 
that was not accessible in the conventional equilibrium DSF measured 
on the ground state. To understand that, we employ the Lehmann's
representation of the nDSF and show that the continua there indeed 
can be associated with the spinon excitations. By inserting the energy
eigenstates $\ket{m}$ and $\ket{n}$ in $S^{\alpha\beta}(k, \omega)$
Eq.~(\ref{Eq:DSF}), we find 
\begin{equation}
\label{Eq:LmRep}
S^{\alpha\beta}(k, \omega) = 2\pi\sum_{m,n}\langle\Psi|m\rangle
\langle m |\sigma_k^\alpha|n\rangle \langle n|\sigma_{-k}^\beta|\Psi\rangle 
\, \delta(\omega - \omega_{nm})
\end{equation}
where $\sigma_{k}^\alpha \equiv \frac{1}{\sqrt{L}}\sum_j e^{- {\rm i}k j} 
\sigma_j^\alpha$. With this representation Eq.~(\ref{Eq:LmRep}), 
we can now understand the peak position
$\omega = \omega_{mn}$ 
in the nDSF --- pole in the Green's function --- indeed represents
the intrinsic excitation of the system, as $\omega_{nm} \equiv 
E_n - E_m$ is the energy difference between the eigenstates 
$\ket{m}$ and $\ket{n}$ of the system. Therefore, different from
the ground state equilibrium DSF where the excitation energy
$\omega$ are positive semi-definite, here $\omega_{mn}$ can be 
both positive and negative, containing both spinon creation and 
annihilation processes that lead to the bowtie and shell continua 
in Figs.~\ref{Fig1},\ref{Fig2}.

With this spectral representation Eq.~(\ref{Eq:LmRep}), the excitation 
continua and distinct dispersions in the nDSF can be understood 
in a generic sense. For instance, in Fig.~\ref{Fig2}(e) we show the nDSF
$S^{zz}_{\rm FMX}(k,\omega)$ where the initial state $\ket{\Phi}_{\rm FMX}$ 
is very close to the nearly polarized state induced by the large
transverse field, and the nDSF clearly reflects the single-particle 
dispersion $\epsilon(k) = 2\sqrt{1 - \sin(2\gamma)\cos k}$.

\subsection{Detecting the QCP via the nDSF} 
As mentioned above, the nDSF results show distinct features in 
the {FM} and QP phases in Fig.~\ref{Fig2}, naturally it suggests 
that the nDSF can be used to accurately locate the QCP. 

In Fig.~\ref{Fig3}(a) we show the nDSF $S^{xx}_{\rm FMZ}(k=0,\omega)$
which reflects the prominent two-spinon continua in the FM phase 
[c.f., Fig.~\ref{Fig2}(a), cut at $k=0$], which changes to the 
spinon-antispinon feature [Fig.~\ref{Fig2}(d), also cut at $k=0$] 
in the QP phase with $\gamma > \gamma_c$. To be specific, 
the behaviors of $S^{xx}_{\rm FMZ}(k=0,\omega)$ change abruptly 
as $\gamma$ exceeds the critical value of $\gamma_c=\pi/4$. 
The excitation continua get significantly weaker and fide out 
in the QP phase, and instead there arise a strong intensity 
at $\omega=0$ which can be ascribed to the shell excitation cut 
at $k=0$ [c.f., Fig.~\ref{Fig2}({d})]. Such a transition of excitation 
behaviors is also reflected in $S^{zz}_{\rm FMX}(k=\pi,\omega)$
in Fig.~\ref{Fig3}(b), where the single-particle excitation distinct in 
the QP phase smears out 
and eventually softens to an excitation 
energy of $\omega=0$ in the $\gamma=0$ Ising limit.

In Fig.~\ref{Fig3}(c) we further show $S^{zz}_{\rm FMZ}(k=0,\omega)$
that hosts a clear static peak in the $\gamma=0$ limit. Such peak gradually 
smears out and becomes branched for $\gamma>\gamma_c$, and the 
splitting that represents the excitation energy $\omega = \omega_{mn}$ 
equals the eigenvalue differences $\pm 2$ in the $\gamma=\pi/2$ limit. 
Similarly, in Fig.~\ref{Fig3}(d) the static peak in $S^{xx}_{\rm FMX}(k=0,\omega)$ 
at $\omega=0$ in the $\gamma>\gamma_c$ side splits into three branches 
in the FM phase with $\gamma<\gamma_c$, forming a triad-like shape. 
According to the results in Fig.~\ref{Fig3}, we find that, although the 
quantum simulator is not initially set at the ground state of the system, 
the dynamical information in the nDSF nevertheless can be used to 
detect the QCP with a clear resolution.

% ============ Fig. 4 =========== %
\begin{figure}[ht]
    \centering
    \includegraphics[width = 1\linewidth]{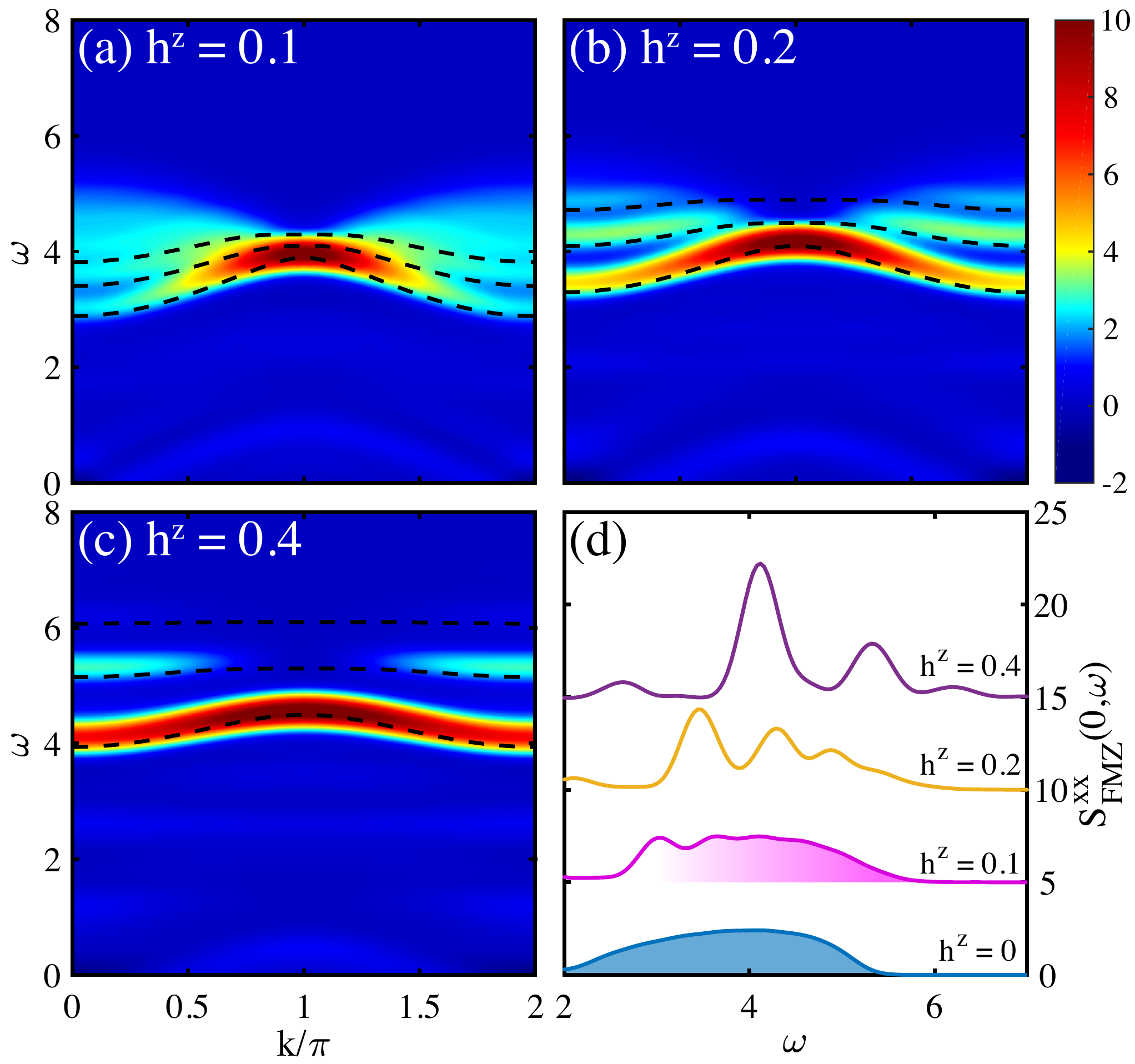}
    \caption{(a-c) show the nDSF $S^{xx}_{\rm FMZ}(k,\omega)$ of the 
    TFI chian in the FM phase ($\gamma = \pi/8$) under longitudinal 
    fields $h^z = 0.1, 0.2, 0.4$, where the spinon bound states and 
    the asymptotically free spinons at higher energies, especially 
    when $h^z$ is small, can be clearly identified. The dispersions 
    of the spinon bound states agree very well with the perturbative 
    calculations (see Appendix Sec.~\ref{App:SpinonBound}), as labeled 
    by the black dashed lines. (d) gives the $k$-cut of (a-c) at the 
    $\Gamma$ point of $k = 0$, the two-spinon continuum of the same 
    $\gamma$ while under a zero longitudinal field is also shown as a 
    benchmark. {The blue shadow regime in (d) represents the spinon 
    {continuum}, while the pink regime at $h^z=0.1$ reflects asymptotically
    free spinons alongside with the low-energy peaks of spinon bound 
    states.} The lines under finite $h^z$ fields in panel (d) are each
    shifted by a constant of $\Delta\omega = 5$ for the sake of clarity.}
    \label{Fig4}    
    \end{figure}
\subsection{Bound States and Asymptotically Free Spinons}
When external longitudinal fields are applied {additionally}, 
the spinons get confined and form spinon bound states. Recall that 
a spinon corresponds a domain wall in the FM phase of TFI model, 
and therefore between two spinons there exists a domain of an opposite 
spin orientation. Such a 1D domain, string-like object, will pose string 
tension between two spinons in the presence of a finite longitudinal field. 
As the confining potential is linear versus the string length {$n$} 
[size of the domain, see Fig.~\ref{Fig1}(e)], i.e., $V(n) \sim {2}h^z \, n$, 
it introduces a linear attractive interaction between a pair of spinons. 
The spinon continua at $h^z=0$ thus split to several leaves with 
{clear lineshapes,} which correspond to the spinon bound states 
with different length $n$ and can be distinguished clearly in the 
nDSF $S^{xx}_{\rm FMZ}(k,\omega)$ in Fig.~\ref{Fig4}(a-c). A 
perturbative calculation reveals that each of them corresponds to 
a bound state with distance $n$ (here up to $n=3$) between two 
constituent spinons (see Appendix Sec.~\ref{App:SpinonBound}). 
Moreover,  when the excitation energy becomes high enough, the spinons 
become very dense and close to each other in real space, thus the effective 
attraction $V(n)$ gradually becomes irrelevant. Such asymptotically free 
spinons are evident in the small field $h^z=0.1$ case, where excitation 
continuum is still present in the higher energy regime, as clearly depicted 
in Fig.~\ref{Fig4}(d).

\subsection{Realization of Dynamical Quantum Simulation with Rydberg Atoms}
The proposed phenomena can be observed on contemporary quantum 
simulation platforms, for instance the Rydberg lattice gases based on arrays 
of optical tweezers \cite{Rydberg51Qubit,GHZLukin2019,256Qubit}, where 
each qubit is defined on the  atomic ground state $\ket{0}$ and one selected 
Rydberg state $\ket{1}$. The lattice constant can be adjusted properly such
that  the intra-atom coupling  mediated by the van der Waals interactions 
between Rydberg atoms together with the Rabi coupling and frequency
detuning can be well approximated as a quantum Ising model with independently tunable transverse field{s} and longitudinal field{s} 
\begin{equation}
    H_I\equiv V\sum_{i}\sigma_i^z\sigma_{i + 1}^z + \Omega\sum_{i} \sigma_{i}^x + \sum_{i}\Delta_i \sigma_{i}^z.
\end{equation}
The many-body system is initialized as $\ket{\Psi_0}\equiv \underset{i=1}{\bigotimes}  \ket{0}_i$. Then the essential step for the experimental 
implementation being the measurement of Eq. \ref{Eq:Correlation} can be 
achieved by applying a sequence of gates/operations to the qubits followed 
by a  projective measurement of all qubits in the $\{ \ket{0}, \ket{1}\}$ basis. 

The sequence of gates/operations are composed of $U, e^{-{\rm i}Ht_2}, \sigma_j^{\beta}, e^{-{\rm i}H(t_1-t_2)}, \sigma_i^{\alpha}, e^{{\rm i}Ht_1}$ and $U^\dag$, where the local unitary operation $U$ rotates the initial state $\ket{\Psi_0}$ 
to a given product state $\ket{\Phi}$ in Eq.~(\ref{Eq:Correlation}). Thus, 
after applying them sequentially to the system one obtains a state 
\begin{equation}
    \ket{\Psi}=U^\dag e^{{\rm i}Ht_1}  \sigma_i^{\alpha} e^{-{\rm i}H(t_1-t_2)} 
\sigma_j^{\beta} e^{-{\rm i}Ht_2} U \ket{\Psi_0}.\end{equation}
One can immediately realize 
that the overlap between $\ket{\Psi}$ and  $\ket{\Psi_0}$ exactly yields the 
unequal-time correlation in Eq. \ref{Eq:Correlation}. In order to implement the
operation $e^{{\rm i}Ht_1}$, although it is difficult to directly reverse the sign of the system Hamiltonian due to the sign of the interaction, one can effectively realize 
the whole unitary operation by sandwiching a revised Hamiltonian $H'$ with 
$\sigma^x$ operations on the odd sites, i.e.,  $\Big(\underset{i \in odd} {\bigotimes} \sigma_{i}^x \Big)H' \Big(\underset{i \in odd} {\bigotimes} \sigma_{i}^x\Big)$, where $H'$ is related to $H_I$  as 
\begin{equation}
    H' =  V\sum_{i}\sigma_i^z\sigma_{i + 1}^z - \Omega\sum_{i} \sigma_{i}^x - \sum_{i\in even}\Delta_i \sigma_{i}^z + \sum_{i\in odd}\Delta_i \sigma_{i}^z.
\end{equation}
One can show by Taylor expansion that 
\begin{equation}
    e^{{\rm i}H_I t}= \Big(\underset{i \in odd} {\bigotimes} \sigma_{i}^x \Big)e^{-{\rm i}H't} \Big(\underset{i \in odd} {\bigotimes} \sigma_{i}^x \Big),
\end{equation}
where the~r.h.s.~is experimentally implementable. 

\section{Discussion and Outlook}
Overall, in this work we explored the non-equlibirum DSF measured 
on the available states in contemporary quantum simulators. We find, 
based on the calculations on a prototypical quantum Ising model system, 
the spin fractionalization can be observed by measuring the nDSF 
on the direct product states and alike.
We show, with solid numerical evidence, that nDSF 
can be employed to detect the deconfined spinons and their rich 
creation/annihilation processes not accessible in solid-state experiments, 
as well as bounded and asymptotically free spinons under finite 
longitudinal fields. Moreover, the QCP can be identified with the nDSF, 
even without preparing the delicate ground state.

With this we advocate that these nDSF results have rich connections 
with the quantum simulation experiments. In distinct to previous 
proposals of dynamical quantum simulations~\cite{Baez2020DQS}, 
here we do not require the simulation to start from the ground state 
or thermal Gibbs state of the Hamiltonian, but from direct product 
states that are readily accessible in quantum simulators. Such 
dynamical quantum simulations can offer valuable insight into the 
quantum many-body systems of interest.

\acknowledgments
We are indebted to Tao Shi, Kai Xu and Peng Xu for 
stimulating discussions. This work is supported by NSFC 
under Grant Nos.~11974036, 11834014, 11904018, and 12047503. 
We thank the high performance computation cluster at ITP-CAS 
for their technical support and generous allocation of CPU time. 

%% ############################## %
%% ######## Appendix ############ %
%% ############################## %

\appendix
\setcounter{figure}{0}
\renewcommand{\thefigure}{A\arabic{figure}}

\section*{Appendix \\ ``Detecting Confined and Deconfined Spinons in 
Dynamical Quantum Simulations"}

\section{Tensor Network Approach for the Dynamical Calculations}
\label{App:TEBD}
Here we provide details of our tensor network algorithm for the calculations
of the time-dependent correlation functions
\begin{equation}
\label{EqS:TdCF}
S^{\alpha\alpha}(i, j, t_1, t_2 = 0) = \left\langle 
 \Psi\right| 
e^{{\rm i}Ht_1}\sigma_i^\alpha e^{-{\rm i}Ht_1} \sigma_j^\alpha \left|\Psi \right\rangle,
\end{equation}
where $\alpha = x \text{ or } z$ denote{s} the two types of spin correlation
functions involved in this work. 

% ========= Fig.A1 ========== % 
\begin{figure*}[ht]
\centering
\includegraphics[width = 0.7\linewidth]{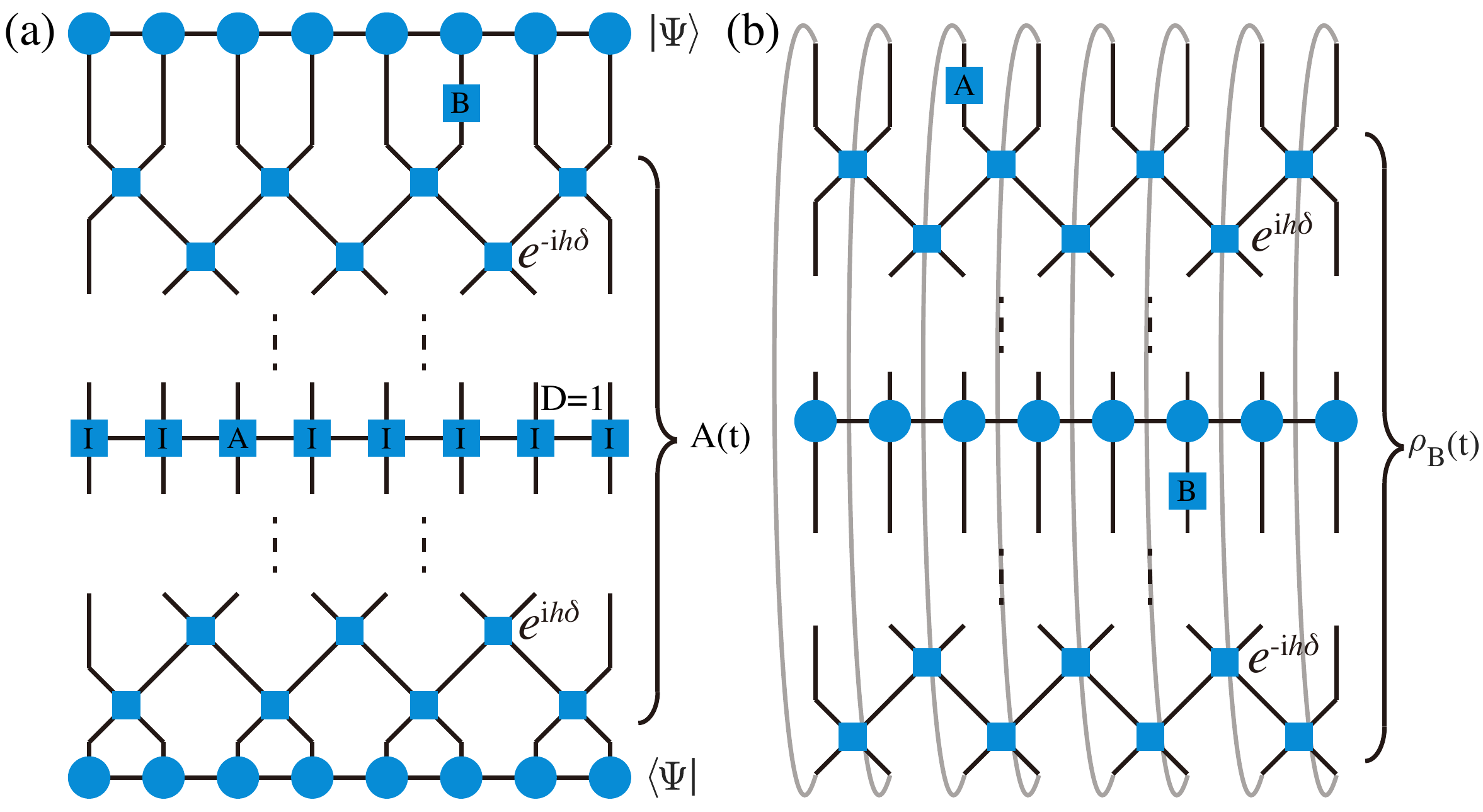}
\caption{Two schemes to contract the tensor network and compute the 
time-dependent correlation function $\left\langle \Psi\right| e^{{\rm i}Ht} 
A e^{-{\rm i}Ht} B\left|\Psi \right\rangle$. (a) follows the Heisenberg 
picture and (b) the Schr{\"o}dinger picture. In both cases, we use 
TEBD technique to evolve and truncate the MPO.}
\label{App_Fig1}    
\end{figure*}
% =========================== % 

% ========= Fig.A2 ========== % 
\begin{figure*}[ht]
\centering
\includegraphics[width = 0.7\linewidth]{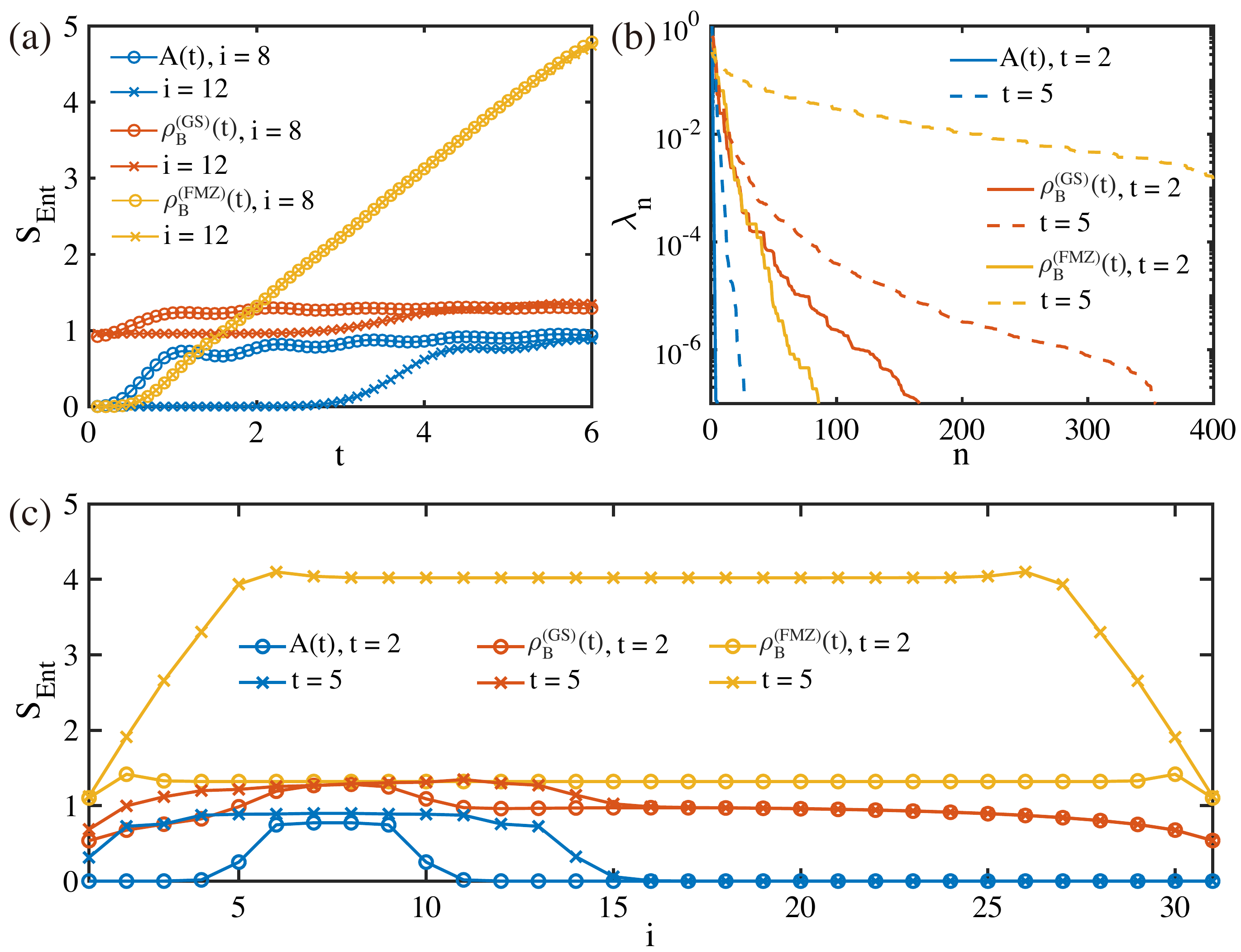}
\caption{(a) shows the entanglement entropy $S_{\rm Ent}$ versus real time 
$t$ of the TFI chain at QCP with $L = 32$, measured at the $i$-th bond. 
Operator $\sigma_8^z$ is chosen in Heisenberg picture, and quenched ground state and FM state along the $\sigma^z$ direction are chosen in Schr{\"o}dinger picture. The entanglement spectra of the $12$-th bond and the spatial distribution of entanglement entropy at $t = 2, 5$ are shown in (b) and (c), respectively.
}
\label{App_Fig2}    
\end{figure*}
% ========================= % 

\subsection{Time Evolution of Matrix Product Operators}
Suppose $A$ and $B$ are
two local operators (in practice $\sigma_i^\alpha$ with $\alpha=x,z$),
and we would like to calculate the correlation
$$\left\langle \Psi\right| e^{{\rm i}Ht}Ae^{-{\rm i}Ht} B\left|\Psi \right\rangle.$$ 
To compute it, we can choose to work
in the Heisenberg or Schr{\"o}dinger picture: In the Heisenberg picture, 
we firstly represent $A$ as a matrix product operator (MPO) with 
bond dimension $D = 1$ because of the locality of the operator. 
Then, we use the time evolution block decimation (TEBD) 
approach \cite{Vidal2004TEBD,White2004tDMRG,Vidal2007iTEBD}
equipped with the folding trick~\cite{Banuls2009} to evolve the MPO 
in real time. The time evolution operators $e^{\pm {\rm i}Ht}$ are decomposed 
via the Trotter-Suzuki decomposition with a Trotter step of $\delta \leq 0.1$
resulting in negligible Trotter error.  After obtaining the  MPO form of 
$A(t) \equiv e^{{\rm i}Ht}Ae^{-{\rm i}Ht}$, we contract the tensor network
shown in Fig.~\ref{App_Fig1}(a) and compute the results of the 
correlation function. 

On the other hand, in Schr{\"o}dinger picture we instead calculate the 
time evolution of the density operators (i.e., folded states). 
The (quenched) density operator is
$$\rho_B \equiv B\left|\Psi\right\rangle\left\langle\Psi\right|,$$ 
and we have
\begin{equation}
\left\langle \Psi\right| e^{{\rm i}Ht}Ae^{-{\rm i}Ht} B\left|\Psi \right\rangle = 
\Tr{e^{-{\rm i}Ht}\rho_B e^{{\rm i}Ht}A}.
\end{equation}
As shown in Fig.~\ref{App_Fig1}(b), to compute the correlation function
we firstly represent $\rho_B$ as an MPO with bond dimension $D = 1$, and
then evolve it from both sides until the whole tensor network is contracted.

The entanglement in either $A(t)$ or $\rho_B(t)$ grows rapidly with time,
hence we can only simulate relatively short real time with high accuracy.
In Fig.~\ref{App_Fig2} we present a comparison between various 
schemes, where the MPO bond dimension is selected as $D = 400$.
From the numerical results we find, at least for TFI chains, the Heisenberg
picture works better, as the accumulated entanglement is significantly 
smaller and thus allows for a longer time evolution.  Moreover, when working in the Schr{\"o}dinger 
picture, one needs to compute $\rho_B(t)$ multiple times for different 
{initial states $|\Phi\rangle$}, which is highly time consuming as compared to 
the Heisenberg picture calculations with $A(t)$. Lastly, for operator $\sigma^x$, 
the MPO form of $\sigma_i^x(t)$ in the TFI chain has a rather compact
representation with bond dimension $D = 4$. As such, we choose the 
Heisenberg picture exclusively.

For the calculations 
in the main text, we eventually choose $D = 800$ and perform the
simulations to a time period of $t_M = 12$, with truncation error  
$\lesssim 10^{-5}$.

\subsection{Bond Inversion and Time Reversal Symmetries}
Note that both the TFI Hamiltonian in Eq.~(1) of the main text, 
even with longitudinal external fields, as well as the density operators $\rho$  
we consider in the present work have a $\mathbb{Z}_2^2$ symmetry, 
with the two 2-nd ordered generators, i.e. bond-inversion $\mathcal{B}$ 
and {the generalized} time reversal $\mathcal{O} \equiv \mathcal{R}_y\mathcal{T}$, 
where $\mathcal{R}_y$ is the $\pi$-rotation along the $\sigma^y$ 
direction and $\mathcal{T}$ represents the time reversal. Assuming 
an even length $L$, we have the following two equations that can be
exploited to accelerate our time-dependent calculations.
\begin{equation}
    \left\{
    \begin{aligned}
        &S^{\alpha\alpha}(L - i + 1, L - j + 1, t) = S^{\alpha\alpha}(i, j, t)\\
        &S^{\alpha\alpha}(i, j, -t) = \overline{S^{\alpha\alpha}(i, j, t)}
    \end{aligned}
    \right.
\end{equation}
With these relations, we can save the computational costs in the sense 
that only about one quarter of correlation functions are needed to be computed,
which greatly facilitates the calculations.

The first equation follows the bond inversion symmetry, and we provide a brief proof of the second equation. Firstly,
\begin{equation}
    \begin{aligned}
    S^{\alpha\alpha}(i, j, -t) =& \Tr{
        \rho e^{-{\rm i}Ht}\sigma_i^\alpha e^{{\rm i}Ht}\sigma_j^\alpha}\\
    =& \Tr{\mathcal{O}\rho e^{{\rm i}Ht}\mathcal{O}\sigma_i^\alpha\mathcal{O} e^{-{\rm i}Ht}\mathcal{O}\sigma_j^\alpha\mathcal{O}^2 },
    \end{aligned}
\end{equation}
where $\mathcal{O}\rho \mathcal{O} = \rho$ and $\mathcal{O} e^{\pm {\rm i}Ht} \mathcal{O} = e^{\mp {\rm i}Ht}$ are used. Considering $\mathcal{O} \sigma^x \mathcal{O} = \sigma^x$, $\mathcal{O} \sigma^y \mathcal{O} = -\sigma^y$, $\mathcal{O} \sigma^z \mathcal{O} = \sigma^z$, no matter $\alpha = x, y$ or $z$, we always have
\begin{equation}
    {\rm R.H.S.} = \Tr{\mathcal{O}\rho e^{{\rm i}Ht}\sigma_i^\alpha e^{-{\rm i}Ht}\sigma_j^\alpha\mathcal{O}}.
\end{equation} 
Note that $\mathcal{O}$ is anti-linear so that $\Tr{\mathcal{O}\cdots\mathcal{O}} = \overline{\Tr{\cdots}}$, finally
\begin{equation}
    S^{\alpha\alpha}(i, j, -t) 
    = \overline{\Tr{\rho e^{{\rm i}Ht}\sigma_i^\alpha e^{-{\rm i}Ht}\sigma_j^\alpha}}
    = \overline{S^{\alpha\alpha}(i, j, t)}.
\end{equation}

Worth to point out, although Ne\'el state along the $\sigma^z$ direction is not 
invariant under either $\mathcal{B}$ or $\mathcal{R}_y\mathcal{T}$, 
it is invariant under $\mathcal{R}_x\mathcal{B}$ and 
$\mathcal{R}_x\mathcal{T}$. Thanks to such symmetry, 
we can also accelerate the calculations in such case.

\subsection{3. Window Functions in Fourier's Transformation}
Now we consider the Fourier's transformation (FT) from time space to frequency space, i.e.,
\begin{equation}
\mathcal{F}[f(t)](\omega) \equiv \int_{-\infty}^\infty {\rm d}t e^{{\rm i}\omega t} f(t),
\end{equation}
which can be computed with the numerical integral. 
In the tensor network calculations, we have only access to a finite time 
$t$ window, i.e., $t\in[-t_M, t_M]$. Hence,  we adopt the Parzen function, 
a Gauss-like window function but with compact support, 
\begin{equation}
    W(x; a) = \left\{
    \begin{aligned}
        &1 - 6\left|\frac{x}{a}\right|^2 + 6\left|\frac{x}{a}\right|^3,\ &|x| \leq \frac{a}{2}\\
        &2\left(1 - \left|\frac{x}{a}\right|\right)^3,\ &\frac{a}{2} < |x| \leq a\\
        &0,\ &|x| > a
    \end{aligned}
    \right.
\end{equation}
and multiply it to the time-dependent correlation functions in order 
to suppress the non-physical oscillating due to time truncation
\cite{Kuhner1999PRB}. Hence, we actually calculate
\begin{equation}
\begin{aligned}
&\int_{-t_M}^{t_M} {\rm d}t e^{{\rm i}\omega t} f(t)W(t; t_M) \\
= &\int_{-\infty}^{\infty} {\rm d}t e^{{\rm i}\omega t} f(t)W(t; t_M)\\ 
= &\frac{1}{2\pi}\int_{-\infty}^\infty {\rm d}\omega^\prime \mathcal{F}[f(t)](\omega^\prime)\mathcal{F}[W(t; t_M)](\omega - \omega^\prime)\\
= &\frac{1}{2\pi}\mathcal{F}[f(t)](\omega)\ast \mathcal{F}[W(t; t_M)](\omega),
\end{aligned}
\end{equation}
where the convolution kernel is
\begin{equation}
    \frac{1}{2\pi}\mathcal{F}[W(t; t_M)] = \frac{96\sin^4\left(t_M\omega/4\right)}{\pi t_M^4\omega^4},
\end{equation}
whose standard error in frequency space is $2\sqrt{3}/t_M$, characterizing the frequency resolution.

\subsection{Linear Prediction Techniques in the Dynamical Calculations}
Linear prediction is a commonly used method to improve the energy resolution in dynamical properties \cite{White2008PRB,Barthel2009}. 
Its key point is use the linear combination of the data we have 
to predict the regime we do not have access. Consider a series 
$\{x_i\}$, $k$-order linear prediction gives
\begin{equation}
    \hat{x}_n = \sum_{i = 1}^k a_i x_{n - i},
\end{equation}
where the coefficients $a_i$ are determined by fitting the known data. 
There are many different fitting algorithms to calculate the coefficients, 
e.g., Burg's method~\cite{kay1988modern} adopted in the present work. 
With this, we can predict the correlation function with $t > t_M$, 
which can significantly improve the frequency resolution. 

We call a series line spectral, if the FT of it is a summation of a few Dirac 
$\delta$-functions, and the general theory of linear prediction guarantees
that when the results are {closer} to a line spectral series, the linear
prediction will work better and provide more accurate results\cite{LP_Book}. 
Due to this consideration, we apply linear prediction to the time-dependent 
correlation functions in momentum space
\begin{equation}
    S^{\alpha\beta}(k, t)\equiv \frac{1}{L}\sum_{i,j}e^{-{\rm i}k(i - j)}
S^{\alpha\beta}(i, j, t).
\end{equation}

\begin{figure*}[ht]
\centering
\includegraphics[width = 0.9\linewidth]{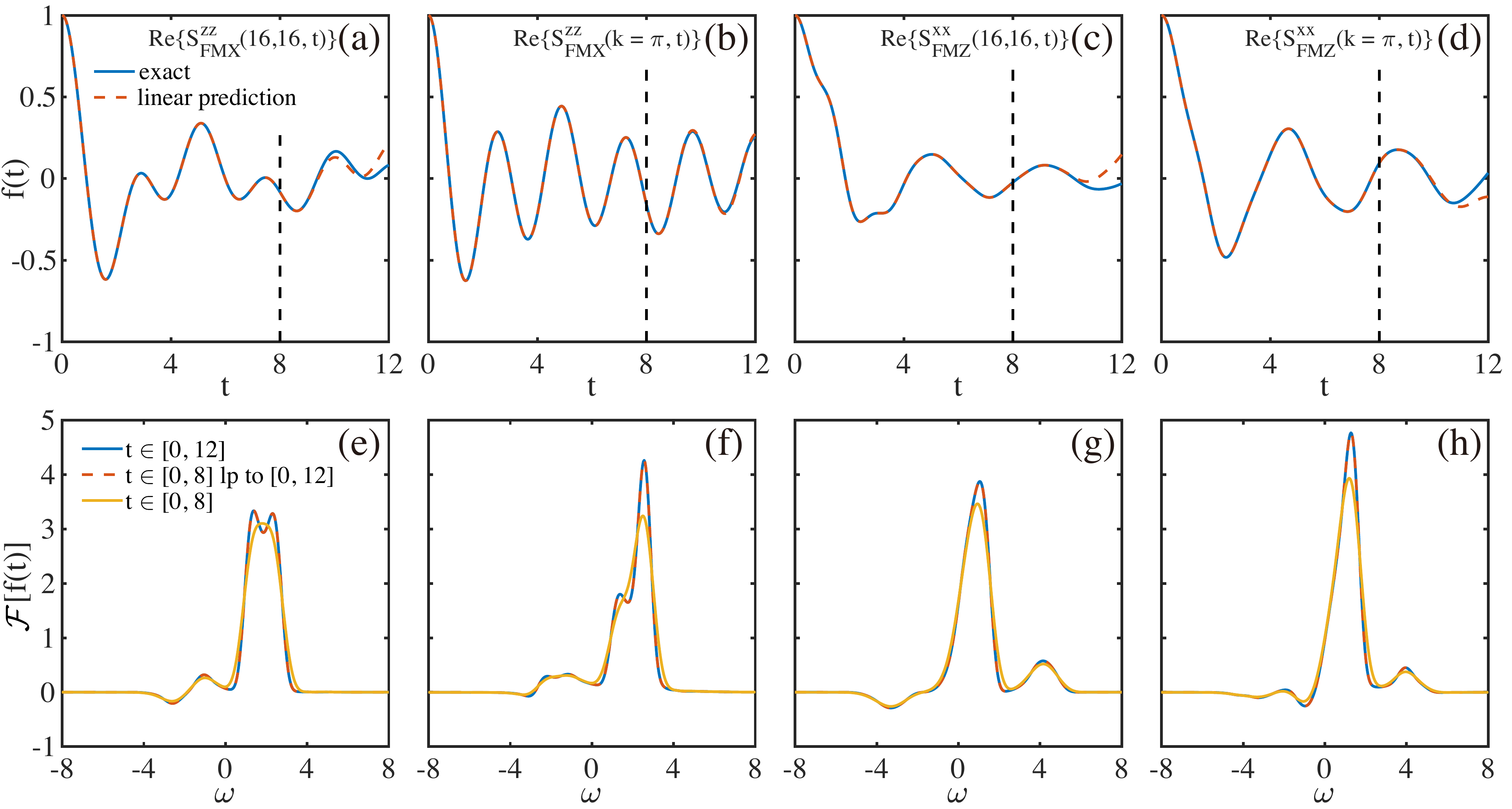}
\caption{Panels (a-d) show the benchmarks of the linear prediction
from $t\in[0, 8]$ to $t\in[0, 12]$ versus the numerically exact results.  
We show the time-dependent correlation in the QP phase 
($\gamma = 3\pi/8)$, measured (a) on the FMX state and 
(c) on the FMZ state. (b,d) are the counterparts of (a,c)
after a spatial FT, and  the lower panels (e-h) show
the FT in frequency space of the time-dependent correlation 
functions in (a-d).}
\label{App_Fig3}    
\end{figure*}

Fig.~\ref{App_Fig3}{(f)} shows the benchmark results on the nDSF 
$S_{\rm FMX}^{zz}(k = \pi, \omega)$, which reflects the magnon 
excitation in the QP phase and is standardly line spectral up to finite size effects. As a result, the linear prediction from 
$t_M = 8$ to $t_M = 12$ works excellently as shown in 
Fig.~\ref{App_Fig3}(b). However, compared to that in momentum 
space, the linear prediction of real-space correlations like $S^{zz}(i, j, t)$ 
results in a visible error as seen in Fig.~\ref{App_Fig3}(a). 
Therefore, it suggests us to work in momentum space. For the spectrum 
with spinon continua, e.g., the correlation $S_{\rm FMZ}^{xx}(k = \pi, \omega)$ 
$k$-cut at $k = \pi$ in Fig.~\ref{Fig2}{(d)} of the main text, the linear 
prediction does not works well in both real and momentum spaces, c.f., 
Fig.~\ref{App_Fig3}(c,d). Nevertheless, the obtained spectral functions 
still coincide with the exact results, as shown in Fig.~\ref{App_Fig3}(g,h), 
which may be due to the fact that window function makes the finite value of 
$t \to t_M$ less important. By these trial calculations, we think that our 
spectra calculated after linear prediction in the main text are reliable. 

% ========= Fig.A3 ========== % 
% =========================== % 

\section{Free Fermion Representation of TFI Chain}
\label{App:ExactTFI}
Hamiltonian (\ref{Eq:TFI}) can be transformed to free fermion 
representation via Jordan-Wigner transformation. Here our convention is 
\begin{equation}
    \left\{
    \begin{aligned}
        & \sigma_i^x = 2c_i^\dagger c_i - 1\\
        & \sigma_i^y = K_i(c_i + c_i^\dagger)\\
        & \sigma_i^z = iK_i(c_i - c_i^\dagger)\\
    \end{aligned}
    \right.
\end{equation}
where 
\begin{equation}
K_i = \prod_{j = 1}^{i - 1}\sigma_j^x = 
\prod_{j = 1}^{i - 1}(2c_j^\dagger c_j - 1)
\end{equation}
represents the Jordan-Wigner string. Using 
$\{\sigma_i^y\pm i\sigma_i^z ,\sigma_i^x\} = 0$, 
one can check the fermion relationship
\begin{equation}
\{c_i, c_j^\dagger\} = \delta_{ij},\ \{c_i, c_j\} = 
\{c_i^\dagger, c_j^\dagger\} = 0.
\end{equation}
Then, the free fermion Hamiltonian of TFI chain is
\begin{equation}
\begin{aligned}
H =& -\cos\gamma\sum_i\left(c_i^\dagger + c_i\right)
\left(c_{i + 1}^\dagger - c_{i + 1}\right)\\
 &- \sin\gamma \sum_i\left(c_i^\dagger c_i - c_i c_i^\dagger\right).
\end{aligned}
\end{equation}
Here we assume the periodic boundary condition for
the free fermion Hamiltonian, which does not alter the 
physics in the thermodynamic limit $L \to \infty$. 
Then, we can apply a spatial FT
\begin{widetext}
\begin{equation}
    f_k = \frac{1}{\sqrt{L}}\sum_{n}e^{-{\rm i}kn}c_n,\ k = -\pi, -\pi + \frac{2\pi}{L}, \cdots, 0, \cdots, \pi - \frac{2\pi}{L}
\end{equation}
to get the BdG Hamiltonian
\begin{equation}
    H = \sum_{k \geq 0} [f_k^\dagger\ f_{-k}]\left[
        \begin{matrix}
            &2\cos\gamma\cos k - 2\sin\gamma & -2i\cos\gamma\sin k\\
            &2i\cos\gamma\sin k & -2\cos\gamma\cos k + 2\sin\gamma   
        \end{matrix}
    \right]\left[
        \begin{matrix}
            f_k\\ f_{-k}^\dagger 
        \end{matrix}
    \right] \equiv \sum_{k \geq 0} \Psi_k^\dagger \widetilde{H}(k)\Psi_k . 
\end{equation}
Diagonalize $\widetilde{H}(k) = V(k)D(k)V^\dagger(k)$, where $D(k) = \mathrm{diag}[\epsilon_k, -\epsilon_k]$, we get 
\end{widetext}
\begin{equation}
    H = \sum_{k \geq 0} \epsilon_k\left(
        \eta_k^\dagger \eta_k - \eta_{-k}\eta_{-k}^\dagger
    \right) = \sum_{k}\epsilon_k\left(
        \eta_k^\dagger\eta_k - \frac{1}{2}
    \right)
\end{equation}
with $[\eta_k^\dagger\ \eta_{-k}] = \Psi_k^\dagger V(k)$.
The dispersion reads
\begin{equation}
    \epsilon(k) = 2\sqrt{1 - \sin(2\gamma)\cos k},
    \label{TFI_dispersion}
\end{equation}
and in the limit of Ising coupling ($\gamma \to 0$) 
or transverse field ($\gamma \to \pi/2$), it becomes
\begin{equation}
    \epsilon(k) \simeq 2 - \sin(2\gamma)\cos k.
    \label{TFI_dispersion_cos}
\end{equation}

% ========= Fig. A4 ======== %
\begin{figure*}[ht]
\centering
\includegraphics[width = 0.8\linewidth]{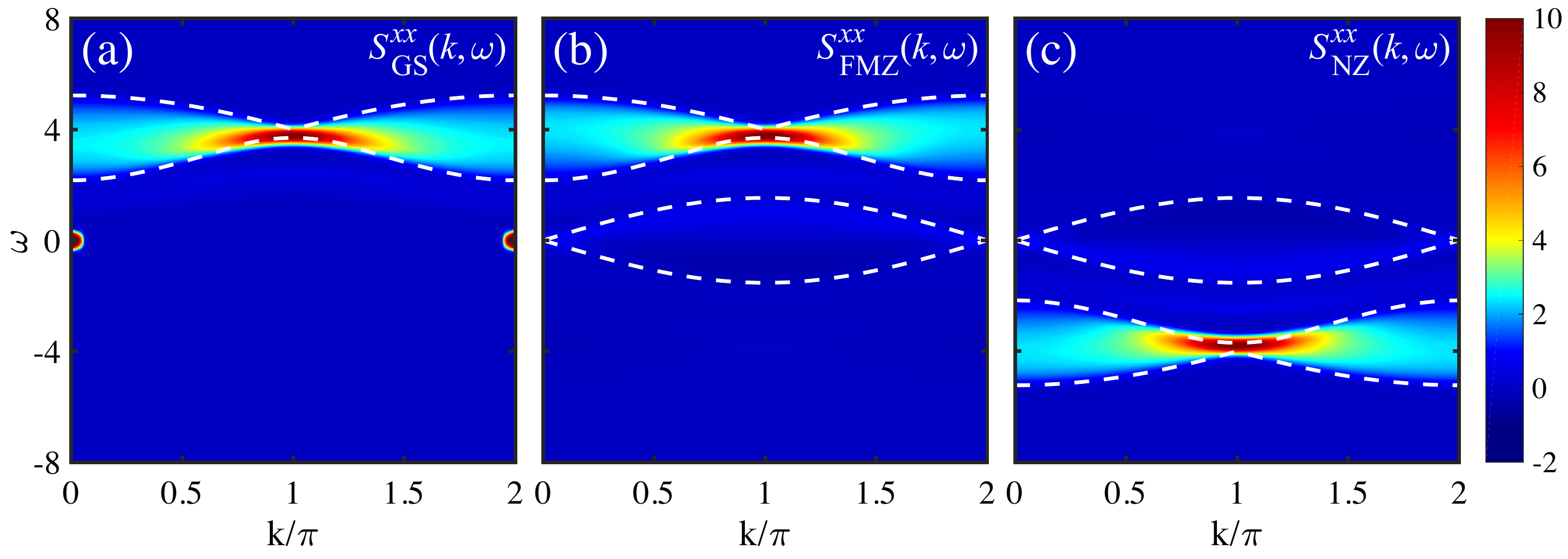}
\caption{The nDSF $S^{xx}(k,\omega)$ of the TFI model
at $\gamma = \pi/8$. (a) is DSF measured on the ground state, 
where only the two-spinon continuum can be identified, except for 
the static peak at the $\Gamma$ point 
due to the finite transverse field.
(b) shows the nDSF measured on $\ket{\Phi_{\rm FMZ}}$, where we 
see both the two-spinon continuum and the weaker spinon-antispinon 
continuum. (c) presents the nDSF measured on the Ne\'el state along the 
$\sigma^z$ direction, which is a product state far from the ground state. 
In this case, the two-antispinon continuum can be clearly recognized.} 
\label{App_Fig4}    
\end{figure*}
% ========================= %

\section{The Bowtie and Shell-like Excitation{s} in the 
nDSF}
\label{App:Bowtie}

\subsection{Excitation Continua in the Gapped Phases}
Here we give a simple derivation of the bowtie and shell-like 
excitation{s} in nDSF. The bowtie excitations
correspond to two-spinon creation, and the shell-like excitations 
correspond to spinon-antispinon process, as shown in 
Fig.~\ref{Fig1} of the main text. Note that in Fig.~\ref{Fig2} 
of the main text we used the exact dispersion Eq.~(\ref{TFI_dispersion}) 
to calculate the boundaries numerically away from the QCP, 
but below we will use the approximative dispersion 
Eq.~(\ref{TFI_dispersion_cos}) so that we can obtain analytical 
results of the boundary. The two-spinon continuum is determined 
by $k = k_1 + k_2$ and $\omega(k) = \epsilon(k_1) + \epsilon(k_2)$. 
Substitute the dispersion $\epsilon(k)$ in Eq.~(\ref{TFI_dispersion_cos}) 
into $\omega(k)$, we have
\begin{eqnarray}
\omega(k)  & =  & 4 - \sin(2\gamma)\cos k_1 - \sin(2\gamma)\cos k_2 \notag \\
& =  & 4 - 2\sin(2\gamma)\cos\left(\frac{k}{2}\right)\cos\left(\frac{k_1 - k_2}{2}\right).
\end{eqnarray}
Then we get the boundary 
$$4 \pm 2\sin(2\gamma)\left|\cos\left(\frac{k}{2}\right)\right|,$$
where we set $\left|\cos\left(\frac{k_1 - k_2}{2}\right)\right| = 1$. 
For the two-spinon annihilation, note that the dispersion of 
antispinon is $-\epsilon(k)$, two-antispinon boundary is just 
$$-4 \pm 2\sin(2\gamma)\left|\cos\left(\frac{k}{2}\right)\right|.$$ 
Similarly, the spinon-antispinon continuum reads
\begin{equation}
\begin{aligned}
    \omega(k) =& \epsilon(k_1) - \epsilon(k_2)\\
    =& -\sin(2\gamma)\cos k_1 + \sin(2\gamma)\cos k_2\\
    =& 2\sin(2\gamma)\sin\left(\frac{k}{2}\right)\sin\left(\frac{k_1 - k_2}{2}\right),
\end{aligned}
\end{equation}
which gives the symmetric upper and lower boundary 
$$\pm 2\sin(2\gamma)\left|\sin\left(\frac{k}{2}\right)\right|.$$ 
The above bowtie and shell-like excitation{s} are shown in 
Fig.~\ref{Fig1}(a) of the main text. 

It is worth to mention that when we use the exact dispersion{,} instead 
of the cos-type one in the Ising or large-field limits{,} the upper and lower
boundaries of the bowtie excitations are no longer symmetric, and there 
will be a finite linewidth at $k = \pi$, as shown in Fig.~\ref{Fig2}(b,d) of
the main text, instead of 
purely a node in Fig.~\ref{Fig1}(a).

\subsection{Excitation Continua at the QCP}
Exactly at QCP, the spinon dispersion reads
\begin{equation}
\label{EqS:QCPSpinon}
    \epsilon(k) = 2\sqrt{2}\left|\sin\left(\frac{k}{2}\right)\right|,
\end{equation}
and the two-spinon continuum can be written as 
\begin{equation}
    \omega(k) = \epsilon(k_1) + \epsilon(k_2) = \epsilon\left(\frac{k}{2} +\delta \right) + \epsilon\left(\frac{k}{2} -\delta \right).
\end{equation}
Firstly we consider the case of $0\leq k < \pi$, then
\begin{equation}
    \omega(k) = \left\{
        \begin{aligned}
            & 4\sqrt{2}\sin\left(\frac{k}{4}\right)\cos\left(\frac{\delta}{2}\right),\ &0 \leq \delta < k/2\\
            & 4\sqrt{2}\cos\left(\frac{k}{4}\right)\sin\left(\frac{\delta}{2}\right),\ &k/2 \leq \delta < 2\pi - k/2\\
            & -4\sqrt{2}\sin\left(\frac{k}{4}\right)\cos\left(\frac{\delta}{2}\right),\ &2\pi - k/2 \leq \delta < 2\pi\\
        \end{aligned}
    \right.
\end{equation}
Considering all these cases, $\omega(k)$ reaches its lower bound $2\sqrt{2}\sin\left(k/2\right)$ at $\delta = k/2$ and its upper bound $4\sqrt{2}\cos\left(k/4\right)$ at $\delta = \pi$. Similarly, in the case of 
$\pi\leq k < 2\pi$, $\omega(k)$ reaches its lower bound $2\sqrt{2}
\sin\left(k/2\right)$ at $\delta = k/2$ and its upper bound $4\sqrt{2}
\sin\left(k/4\right)$ at $\delta = 0$. 
Also consider the case of $0\leq k < \pi$, the spinon-antispinon 
continuum can be written as 
\begin{equation}
    \omega(k) = \epsilon(k_1) - \epsilon(k_2) = \epsilon\left(\frac{k}{2} +\delta \right) -\epsilon\left(\frac{k}{2} -\delta \right).
\end{equation}
Note that it is an odd function of $\delta$, the boundary of 
spinon-antispinon continuum is much easier to calculate. 
Because of the sine-type dispersion (\ref{EqS:QCPSpinon}), 
$\partial\omega/\partial \delta \geq 0$ holds for all $|\delta| < k/2$ 
and $\partial\omega/\partial \delta \leq 0$ holds for all 
$k/2 < |\delta| < \pi$. Hence, $\omega(k)$ must reach its bound 
at $\delta = \pm k/2$ or $\pm \pi$. However, the latter must result in 
$\omega(k) = 0$ because it is both odd and $2\pi$-periodic. Finally, 
we get its boundary $\pm 2\sqrt{2}\left|\sin\left(k/2\right)\right|$. 
The bowtie and shell excitations at QCP are illustrated 
in Fig.~\ref{Fig1}(b) of the main text.

\subsection{Two-Spinon Annihilation Continua}
\label{App:SubNegBowtie}

We have shown the two-spinon and spinon-antispinon continua 
in the nDSF measured on direct product state{s} in the main text. 
The two-spinon annihilation process can also be seen if we choose 
states with high energy, i.e., far from the ground state. Fig.~\ref{App_Fig4} shows the nDSF $S^{xx}(k,\omega)$ at $\gamma = \pi/8$ measured on
different product states, equilibrium DSF measured on ground state is 
also shown in panel (a) as a comparison. 

\begin{widetext}
\section{Dispersion of Spinon Bound States of TFI Chain with 
Longitudinal Fields}
\label{App:SpinonBound}

The Hamiltonian of TFI chain under a longitudinal field reads
\begin{equation}
H = -\cos{\gamma}\sum_{i} \sigma_i^z\sigma_{i + 1}^z - 
\sin{\gamma} \sum_{i} \sigma_{i}^x - h^z\sum_{i} \sigma^z,
\end{equation}
where $h^z$ is the strength of the longitudinal field.
We consider the FM phase, and take the transverse fields 
as perturbation in the $\gamma \to 0$ limit. Denote 
\begin{equation}
|i; n\rangle \equiv |\uparrow \cdots \uparrow \underbrace{\downarrow_{i} 
\cdots \downarrow_{i + n - 1}}_{n}\uparrow \cdots\rangle 
\end{equation}
as the state with two spinons, where $i$ is the starting site and $n$ is 
the length of the domain. We also assume $h^z \ll 1$ such that the Ising 
couplings dominate the Hamiltonian, hence the states with more spinons 
can be neglected because of the large gap $\Delta \gtrsim 2\cos\gamma$. 
All such states $\{ \ket{i;n} \}$ span a subspace, and we perform the first
ordered degenerate perturbation calculations in this subspace. After shifting
zero point of energy s.t. $\langle \uparrow\cdots\uparrow|H|\uparrow\cdots\uparrow\rangle = 0$ for convenience,
\begin{equation}
H_{\rm eff}|i; n\rangle = \left(4\cos\gamma + 2nh^z\right)|i;n\rangle 
- \sin\gamma\left(|i; n+1\rangle + |i; n-1\rangle + |i - 1; n+1\rangle + 
|i + 1; n - 1\rangle\right).
\end{equation}
A given bound state is the state with two spinons carrying the same quasi-momentum, hence $n$ is unchanging. So we apply FT for each fixed $n$
\begin{equation}
|k;n\rangle \equiv \frac{1}{L}\sum_j e^{-{\rm i}kj}|j;n\rangle.
\end{equation}
Then, in the Fourier basis the effective Hamiltonian is partially decoupled, 
\begin{equation}
\begin{aligned}
H_{\rm eff}|k; n\rangle = \left(4\cos\gamma + 2nh^z\right)|k;n\rangle
- \sin\gamma\left(|k; n+1\rangle + |k; n - 1\rangle + e^{-ik}|k; n + 1\rangle 
+ e^{ik}|k; n - 1\rangle\right).
\end{aligned}
\end{equation}
Now we fix $k$ and get the effective Hamiltonian
\begin{equation}
\langle k;m|H_{\rm eff}|k;n\rangle = \left(4\cos\gamma + 2nh^z\right)\delta_{mn}
- \sin\gamma \left(1 + e^{-ik}\right)\delta_{m, n+1} - \sin\gamma \left(1 + e^{ik}\right)\delta_{m, n-1}, 
\end{equation}
where the position $m,n \geq 1$. We introduce a cutoff $m,n \leq N$,
which is quite nature as the spinon confinement induced by $h^z$ fields. 
In this {basis}, $H_{\rm eff}$ forms a tridiagonial $N\times N$ Hermitian 
matrix, which can be diagonalized and there are $N$ eigenvalues. Apply 
{this procedure} for each $k$ and we get the dispersion of the $N$ 
spinon bound states.  
\end{widetext}

% ========= %
% References
% ========= %
\bibliography{./QSimDynamicsRefs.bib}

\end{document}